# DISCUSSION ABOUT DIFFERENT METHODS FOR INTRODUCING THE TURBULENT BOUNDARY LAYER EXCITATION IN VIBROACOUSTIC MODELS

Laurent Maxit, Marion Berton, Christian Audoly, Daniel Juvé

**Abstract**
For controlling the noise radiated from vibrating structures excited by turbulent boundary layer (TBL) it is relevant to develop numerical tools for understanding how the structure reacts to TBL excitation. Usually, the wall pressure fluctuations of the TBL are described through statistical quantities (i.e. space-frequency or wavenumber-frequency spectra) which depend on the TBL parameters. On the other hand, the vibro-acoustic models (i.e. Finite Elements, Boundary Elements, Transfer Matrix Methods, Analytical models, etc) evaluate deterministic transfer functions which characterise the response of the considered structures. The first part of this paper focuses on the coupling between the stochastic TBL and the deterministic vibro-acoustic models. Five techniques are presented. Numerical applications on an academic marine test case are proposed in order to discuss the calculation parameters and the interests / drawbacks of each technique. In the second part of the paper, the high frequency modelling with the Statistical Energy Analysis (SEA) method is considered. The focus is placed on the estimation of an important input of this method: the injected power by the TBL into the structure for each third octave band.

Laurent Maxit
Laboratoire Vibrations- Acoustique, INSA de Lyon, Bât. St. Exupéry, 25 bis av. Jean Capelle, 69621 Villeurbanne Cedex ; e-mail : laurent.maxit@insa-lyon.fr

Marion Berton
Centre Acoustique, LMFA UMR 5509, Ecole Centrale de Lyon, 36 av. Guy de Collongue, 69134 Ecully Cedex ; e-mail : marion.berton@ec-lyon.fr

Christian Audoly
DCNS Research, Rond-point de l'Artillerie de la Marine, 83000 Toulon
email: christian.audoly@dcnsgroup.com

Daniel Juvé
Centre Acoustique, LMFA UMR 5509, Ecole Centrale de Lyon, 36 av. Guy de Collongue, 69134 Ecully Cedex ; e-mail : daniel.juve@ec-lyon.fr

# 1.  Introduction

Structures excited by the turbulent boundary layer (TBL) are very common in practical applications. Car, airplanes, trains, and submarines may be excited by pressure fluctuations due to the turbulent flow induced by their motions. In order to reduce the noise radiated from these structures, it is important to understand at the design stage how the structure reacts to the TBL excitation. It is then necessary to develop numerical tools allowing predicting the vibration or the radiated pressure from the structure excited by the turbulent flow. Usually, the calculation process is decomposed in 3 steps:

1 - A stationary hydrodynamic model is used to estimation the TBL parameters over the surface of the structure from its geometry and the flow conditions;

2 - The spectrum of the wall pressure fluctuations is evaluated from the TBL parameters estimated in the previous step and by using one of the models proposed in the literature. Some of them are expressed in the space - frequency domain (like the famous Corcos model [1]) whereas as others are expressed in the wavenumber - frequency domain (like the no less famous Chase model [2]); Discussion about different models and comparison with experiment can be found in [3, 4] for the frequency auto spectrum and in [5, 6] for the normalized wavenumber cross spectrum;

3 - The last step consists in using a vibro-acoustic model to estimate the response of the structure to the pressure fluctuations. The choice of the model depends on the frequency range of interest:

- For the low frequencies, deterministic models considering harmonic excitations are generally considered. For example, it can be a standard Finite Element Model (FEM) for a structural problem or FEM coupled with a Boundary Element Model (BEM) for an acoustic radiation problem. The coupling between the statistical model used to describe the wall pressure fluctuations and the deterministic vibroacoustic model constitute a difficulty in the calculation process described above (i.e. the transition from step 2 to step 3). This topic is specifically addressed in the first part of this paper. Five approaches will be proposed and discussed in Sec. 3 after having recalled the mathematical formulation of the problem in Sec. 2.

- For high frequencies, the Statistical Energy Analysis (SEA) method [7] is generally used to represent the vibro-acoustic behavior of complex structures. As the excitation is characterized in SEA by its time-averaged injected power for each frequency band, it is necessary to evaluate this quantity when considering the TBL excitation. We propose and discuss in Sec. 4 a formula allowing estimating the injected power from the wall pressure spectrum expressed in the wavenumber-frequency space. A methodology is also proposed to take the spatial variations of the TBL parameters into account.

## 2. Vibrating structures excited by random pressure fluctuations

### *2.1 Presentation of the problem*

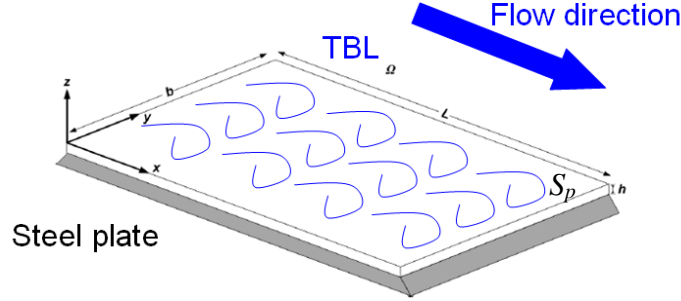

Figure 1. Baffled simply supported plate excited by a homogeneous and stationary TBL.

Let us consider a baffle panel of surface $S_p$ excited by a TBL as shown in Fig. 1. Three assumptions are considered:
- The TBL is assumed to be fully developed, stationary, and homogeneous over $S_p$;
- The plate and the boundary layer are supposed weakly coupled. It is then supposed that the vibration of the plate does not modify the TBL wall pressure excitation. Spectra of the wall pressures over a rigid surface can then be considered;
- It is assumed that the propagation of the acoustic waves into the fluid is not affected by the turbulent flow. Moreover, for the marine applications (i.e. low Mach number), we could also neglect the convective effect on the acoustic wave propagation.

The marine test case considered for the numerical application is composed of a thin rectangular plate simply supported along its four edges and immersed in water on one-side. The flow direction is parallel to the longest edges of the plate (i.e. about x-axis). Numerical values of the physical parameters considered for this test case are given on Tab. 1.

The parameters characterizing the turbulent boundary layer are supposed to be known: $U_\infty$ is the flow velocity, $U_c$, the convection velocity, $\delta$ is the boundary layer thickness, and $\tau_w$ the wall shear stress. From these parameters and the wall pressure models proposed in the literature [2-5], we can define the spectrum of the wall pressure fluctuations acting on the plate. The auto spectrum density of the wall pressure $S_{pp}(\omega)$ is evaluated here considering Goody's model [8] whereas the normalised cross spectrum density is evaluated using the Corcos's model [2]. The later is considered for it simplicity because it provides an analytical expression of the cross spectrum both in the space-frequency domain $\bar{\phi}_{pp}(\zeta,\omega)$ and in the wavenumber-frequency domain $\bar{\phi}_{pp}(k,\omega)$, both.

The spectrum of the wall pressure fluctuations is then given by
- in the physical space $\zeta = (\zeta_x, \zeta_y)$ (i.e. spatial separation):

$$S_{pp}^{TBL}(\zeta,\omega) = S_{pp}(\omega)\bar{\phi}_{pp}(\zeta,\omega), \text{ and,} \qquad (1)$$

- in the wavenumber space $k = (k_x, k_y)$:

$$\phi_{pp}^{TBL}(k,\omega) = S_{pp}(\omega) \bar{\phi}_{pp}(k,\omega). \tag{2}$$

The frequency band of interest is fixed here to [10 Hz – 1 kHz] and is above the hydrodynamic coincidence frequency. It results that the wavelength associated to the convection velocity $\lambda_c = 2\pi U_c / \omega$ is always smaller than the flexural wavelength of the plate.

The objective of the present paper is to estimate the panel response induced by the wall pressure fluctuation defined by its spectrum. In the next section, we give the outlines of the formulation which is described in details in the literature [9-11].

| Parameters | Numerical value |
|---|---|
| Flow velocity | $U_\infty$ =7 m/s |
| Convection velocity | $U_c$ = 5 m/s |
| Boundary layer thickness | $\delta$ = 9.1 cm |
| Wall shear stress | $\tau_w$ = 2.52 Pa |
| Corcos' parameters | $\alpha$ =0.11 ; $\beta$ =0.77 |
| Panel thickness | $h$=1 mm |
| Panel length in the streamwise direction | $L$ =0.455 m |
| Panel length in the crosswise direction | $b$ =0.375 m |
| Panel Young's modulus | $E$=2.1 $\times 10^{11}$ Pa |
| Panel Poisson's ratio | $\nu$ =0.3 |
| Panel mass density | $\rho$ = 7800 kg/m$^3$ |
| Panel damping loss factor | $\eta$ =0.01 |
| Fluid sound speed | $c_0$ =1500 m/s |
| Fluid mass density | $\rho_0$ = 1000 kg/m$^3$ |

Table 1. Physical parameters of the marine test case.

## 2.2 *Mathematical formulation*

$p_b(x,t)$ represents the wall-pressure fluctuations due to the TBL on the plate at point $x$ as a function of time $t$. The plate velocity at point $x$ due to wall-pressure fluctuations, $v(x,t)$ can be expressed as the convolution product

$$v(x,t) = \int_{S_p} \int_{-\infty}^{+\infty} h_v(x,\tilde{x},t-\tau) p_b(\tilde{x},\tau) d\tau d\tilde{x}, \tag{3}$$

where $h_v(x,\tilde{x},t)$ is the velocity impulse response at point $x$ for a normal unit force at point $\tilde{x}$. The improper integral corresponds to the convolution product between the impulse response $h_v(x,\tilde{x},t)$ and the force $p_b(\tilde{x},\tau) d\tilde{x}$ exerted on an elementary surface $d\tilde{x}$ and it gives the plate velocity at point $x$ due to this force (for a time-invariant system). The surface integral over

$S_p$ corresponds to the summation of the effect of the elementary forces over the plate surface and it gives $v(x,t)$ (based on the principle of superposition for a linear system).

As the turbulent flow produces random fluctuations, the plate response is characterised by the auto-correlation function of the velocity, $R_{vv}$. Supposing that the process is stationary and ergodic (i.e. expectation replaced by the limit of a time average), $R_{vv}$ can be written as:

$$R_{vv}(x,t) = \lim_{T \to \infty} \frac{1}{T} \int_{-T/2}^{T/2} v(x,\tau) v(x,t+\tau) d\tau. \tag{4}$$

The Auto Spectrum Density (ASD) of the velocity at point $x$ is defined as the time Fourier transform of $R_{vv}$:

$$S_{vv}(x,\omega) = \int_{-\infty}^{+\infty} R_{vv}(x,t) e^{-j\omega t} dt, \quad \forall \omega \in \mathbb{R}. \tag{5}$$

The same definition is used for the ASD of the wall-pressure fluctuations, $S_{pp}^{CLT}(\zeta,\omega)$.

Note that:

- the Fourier transform $\tilde{f}(\omega)$ of a function $f(t)$, is defined as $\tilde{f}(\omega) = \int_{-\infty}^{\infty} f(t) e^{-j\omega t} dt$ whereas others conventions can be used (for example $\tilde{f}(\omega) = \frac{1}{2\pi} \int_{-\infty}^{\infty} f(t) e^{-j\omega t} dt$). A special attention should be given on this point when the ASD of the wall-pressure fluctuations is extracted of the literature;

- Moreover, the $S_{pp}^{CLT}(\zeta,\omega)$ is here a double-sided spectrum and is a function of the angular frequency $\omega$. The relation with a single-sided spectrum $\bar{S}_{pp}^{CLT}(\zeta,f)$ expressed as a function of only the positive frequency $f$ is $S_{pp}^{CLT}(\zeta,\omega) = (4\pi)^{-1} \bar{S}_{pp}^{CLT}(\zeta,f)$.

Introducing (3) in (4), and the result in (5), we obtain after some manipulations of integrals:

$$S_{vv}(x,\omega) = \iint_{S_p S_p} H_v(x,\tilde{x},\omega) S_{pp}^{TBL}(\tilde{x}-\tilde{\tilde{x}},\omega) H_v(x,\tilde{\tilde{x}},\omega) d\tilde{x} d\tilde{\tilde{x}}, \tag{6}$$

where $H_v(x,\tilde{x},\omega) = \int_{-\infty}^{+\infty} h_v(x,\tilde{x},t) e^{-j\omega t} dt$ is the Frequency Response Function (FRF) in velocity at point $x$ for a normal force at point $\tilde{x}$.

In the same manner, we can obtain the ASD of the radiated pressure at point $z$ into the fluid

$$S_{pp}(z,\omega) = \iint_{S_p S_p} H_p(z,\tilde{x},\omega) S_{pp}^{TBL}(\tilde{x}-\tilde{\tilde{x}},\omega) H_p(z,\tilde{\tilde{x}},\omega) d\tilde{x} d\tilde{\tilde{x}}, \tag{7}$$

where $H_p(z,\tilde{x},\omega)$ is the Frequency Response Function (FRF) in pressure at point $z$ for a normal force at point $\tilde{x}$.

These two equations are the starting point of the following techniques for coupling a wall pressure model with a deterministic vibroacoustic model. In the next section, five different techniques are presented to estimate the vibration response of the panel from (6). These techniques are also applicable to estimate the radiated pressure into the fluid from (7).

## 3. Different approaches to couple a stochastic wall pressure field to a deterministic vibroacoustic model

### 3.1 *Preamble: calculation of Frequency Response Functions*

Different vibroacoustic models can be used to estimate the Frequency Response Functions (FRF) of complex panels radiated into a fluid:
- FEM using Perfectly Matched Layers (PMLs) [12];
- FEM coupled with BEM [13];
- FEM coupled with Infinite Elements [14];
- Transfer Matrix Method (TMM) for infinite multi-layers panels [15];
- Etc…

In these models, different types of harmonic excitations can be considered:
- A normal point force as illustrated on Fig. 2 for estimating a point to point FRF;
- A wall plane wave excitation;
- A specified pressure field over the panel surface.
- Etc…

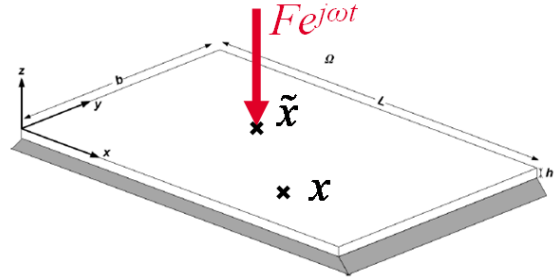

Figure 2. Illustration of the problem for evaluating the FRF between point $x$ and $\tilde{x}$ :
$H_{vv}(x,\tilde{x},\omega) = V_x / F$ .

Basically, for an angular frequency $\omega$ , the equations of motions of the vibroacoustic problem can be written in the matrix form:
$$\mathrm{Dx} = \mathrm{F}, \tag{8}$$
where $\mathrm{D}$ is the dynamic stiffness matrix; $\mathrm{x}$ , the response vector; and $\mathrm{F}$ , the force vector of the considered load case.

The response vector is obtained by inverting the dynamic stiffness matrix:
$$\mathrm{x} = \mathrm{D}^{-1}\mathrm{F}. \tag{9}$$

The FRF can then been determined by extracting the appropriate information in the response vector. In order to simulate the effect of the TBL excitation, many FRFs should be calculated with the vibroacoustic model, and consequently, many load cases should be considered for the process described above (i.e. Eq. (8) and (9)). The management of multi load cases is then an important issue when dealing with TBL excitation. For example, it is generally more efficient to multiply $D^{-1}$ by a force matrix containing the different load cases (i.e. matrix − matrix product) than to achieve a loop over the different load cases and to multiply $D^{-1}$ by the force vector of each considered load case (i.e. loop + matrix − vector product). Moreover, in some situations, for example when using of commercial software, it is not always possible to have this optimal management of the multi load cases. This is why in the following, we will not only indicate the computing time observed on the present test case, but we also indicate the number of considered load cases.

For this present test case, the FRFs have been evaluated using an in-house code based on the PTF (Path Transfer Function) approach ([16-18]). It allows us to have an optimal management of the multi load case under the MATLAB environment. This substructuring method consists in decomposing our problem in two parts: the panel and the semi-infinite fluid. The coupling surface is divided into patches which sizes depend on the considered wavelengths. Each part is characterised separately by PTFs (i.e. Path mobilities for the panel using the modal expansion method, path impedances for the fluid using the Rayleigh integral). Writing the continuity conditions at the coupling interface allows us to assemble the two parts. The particularity of the present model compared to [17, 18] is that the fluid added mass effect is taken into account through the "wet modal frequencies" (which are estimated by assuming the fluid incompressible) instead of using the imaginary part of the acoustic impedance of the fluid domain. This permits to overcome the convergence issue evoked in [18] concerning the patch size criterion. Here, a patch size lower than half the flexural wavelength gives results with good numerical convergence. The numerical process based on the PTF approach has been validated for the test case considered by comparison with results published in the literature [19]. We do not describe more in details these calculations which are out of the scope of the present paper.

## 3.2 *The spatial method*

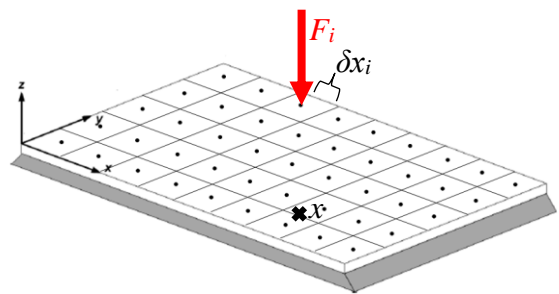

Figure 3. Illustration of the spatial discretization of the panel surface.

The first method for coupling the wall pressure spectrum and the FRFs calculated with a vibroacoustic model is simply based on a regular spatial discretization of the panel surface as shown on Fig. 3. Eq. (6) becomes:

$$S_{vv}(x,\omega) = \sum_{i=1}^{\Theta}\sum_{j=1}^{\Theta} H_v(x,x_i,\omega) S_{pp}^{TBL}(x_i - x_j, \omega) H_v(x, x_j, \omega) \delta x_i \delta x_j, \qquad (10)$$

where $\Theta$ is the number of discrete points and $\delta x_i$ is the elementary surface attributed at the discrete point *i*.

Eq. (10) can be rewritten in the matrix form:
$$S_{vv} = {}^t H S_{pp}^{TBL} H, \qquad (11)$$
with
$$S_{pp}^{TBL} = \begin{bmatrix} \ddots & \vdots & \cdot^{\cdot^{\cdot}} \\ \cdots & S_{pp}^{TBL}(x_i - x_j, \omega) & \cdots \\ \cdot_{\cdot_{\cdot}} & \vdots & \ddots \end{bmatrix}_{\Theta \times \Theta}, \quad H = \begin{bmatrix} \vdots \\ H_v(x, x_i, \omega) \delta x_i \\ \vdots \end{bmatrix}_{\Theta \times 1}. \qquad (12)$$

The point-to-point FRFs, $H_v(x, x_i, \omega)$, $\forall i \in [\![1, \Theta]\!]$, should be evaluated using the vibroacoustic model. The number of load cases corresponds to the number of discrete points, $\Theta$.

A key parameter of this method is the spatial resolution of the discretisation. Results for the test case are plotted in Fig. 4 with different resolutions given as a function on the convective wavelength $\lambda_C$ (which depends on the frequency). The coarser mesh (i.e. $\delta = \lambda_C$) gives poor results excepted at low frequency. It does not allow representing correctly the convective part of the pressure fluctuations. A spatial resolution corresponding to one third of the convective wavelength seems to be a good compromise between the results accuracy and the computing times. Even if the spanwise turbulence wall pressure correlation length of the Corcos model is lower than the convective wavelength (and the streamwise correlation length), a parametric survey shows us that the use of the same criterion for the spatial resolution in the spanwise direction than in the streamwise direction gives relevant results.

We emphasize that the calculation process (based on the matrix form (11)) requires high memory capacity, in particular to store the wall pressure CSD matrix $S_{pp}^{TBL}$. This is why the calculations have not been performed above 500 Hz with our computer (although 1 kHz was initially expected).

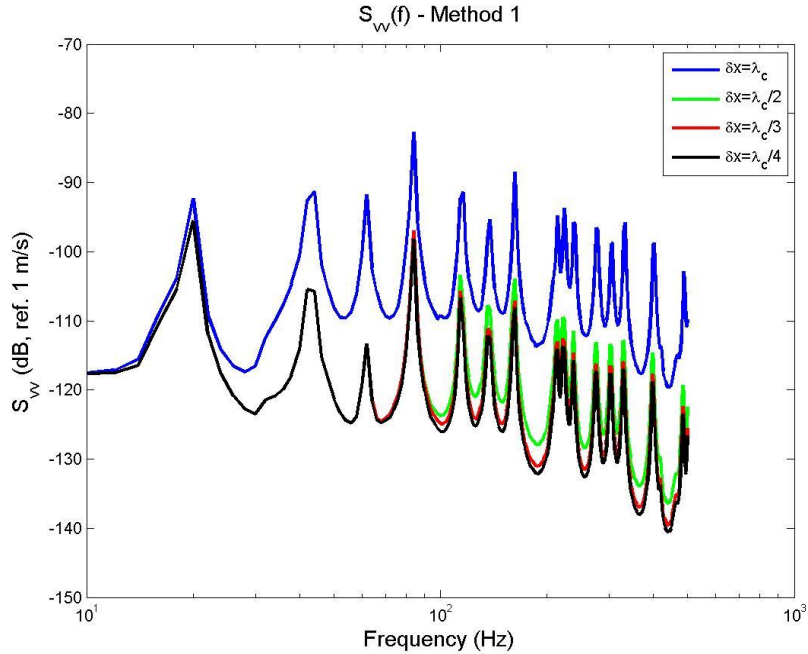

Figure 4. Velocity ASD at point $x$=(0.05, 0.18) for different spatial resolutions: blue, $\delta = \lambda_C$; green, $\delta = \lambda_C/2$; red, $\delta = \lambda_C/3$; black, $\delta = \lambda_C/4$;

### 3.3 *The Choslesky method*

The second method is based on a Choslesky decomposition of the wall pressure CSD matrix $S_{pp}^{TBL}$ [20, 21]:

$$S_{pp}^{TBL}(\omega) = \begin{bmatrix} \ddots & \vdots & \cdot^{\cdot^{\cdot}} \\ \cdots & S_{pp}^{TBL}(x_i - x_j, \omega) & \cdots \\ \cdot^{\cdot^{\cdot}} & \vdots & \ddots \end{bmatrix} = L(\omega) L^T(\omega), \qquad (13)$$

where $L(\omega)$ is a lower-triangular matrix of dimensions $\Theta \times \Theta$ and superscript *T* indicates the transpose of the matrix.

Iin a first step, the method consists in achieving different realizations of the stochastic field characterized by $S_{pp}^{TBL}(\omega)$. The wall pressure vector of the $k^{th}$ realization [21], $p^k$ is given by,

$$p^k(\omega) = L(\omega) e^{j\varphi^k}, \qquad (14)$$

where $\varphi^k$ is a phase vector of $\Theta$ random values uniformly distributed in $[0, 2\pi]$.

So, an ensemble average over a set of realizations of the pressure field approximates the wall pressure CSD matrix:

$$S_{pp}^{TBL}(\omega) \approx E\left[ p^k(\omega) \bar{p}^k(\omega) \right], \qquad (15)$$

where the bar over the complex value indicates the complex conjugate.

In a second step, the vibroacoustic model is used to estimate $v^k(x, \omega)$, the panel velocity at point *x* when the panel is excited by the pressure field, $p^k(\omega)$. This calculation is achieved for a given number of realizations, *K*. The number of load cases considered in the vibroacoustic simulations corresponds then to the number of realization.

Finally, the ASD of the velocity at point *x* is estimated by an ensemble average of the velocity responses, $v^k(x, \omega)$ :

$$S_{vv}(x, \omega) \approx E\left[ v^k(x, \omega) \bar{v}^k(x, \omega) \right]. \qquad (16)$$

We illustrate this approach on the present test case. The velocity responses $v^k(x, \omega)$ of 20 realizations are plotted in grey in Fig. 5. A large dispersion of these responses can be observed. The ensemble average over these 20 realizations (Eq. (16)) is plotted with a black curve on Fig. 5 and compared with the result of the first method on Fig. 6. We can observe a good agreement between the two calculations even when only 20 realizations have been considered. With this approach, the number of load cases is then relatively small.

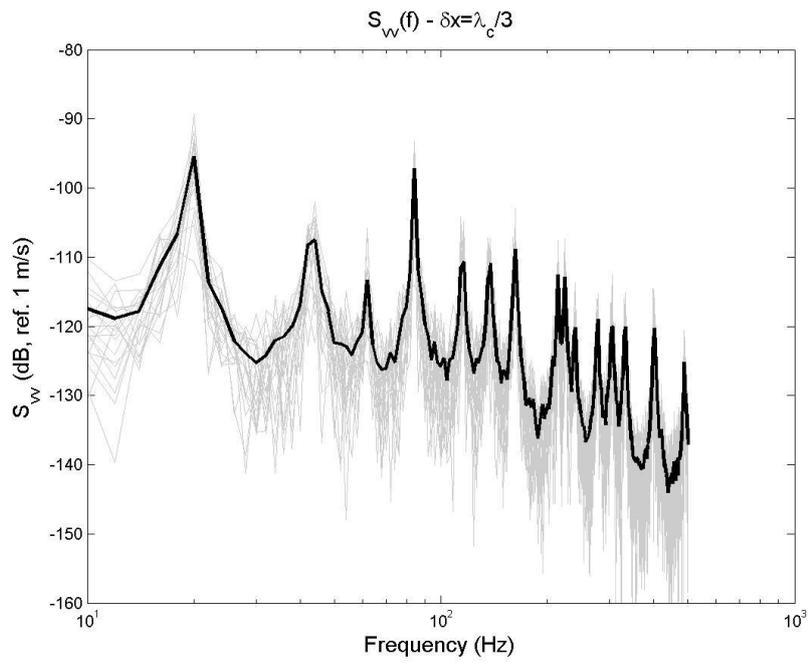

Figure 5. Velocity ASD at point *x*=(0.05, 0.18).
Grey, Results of 20 realizations; Black, Average over the 20 realizations.

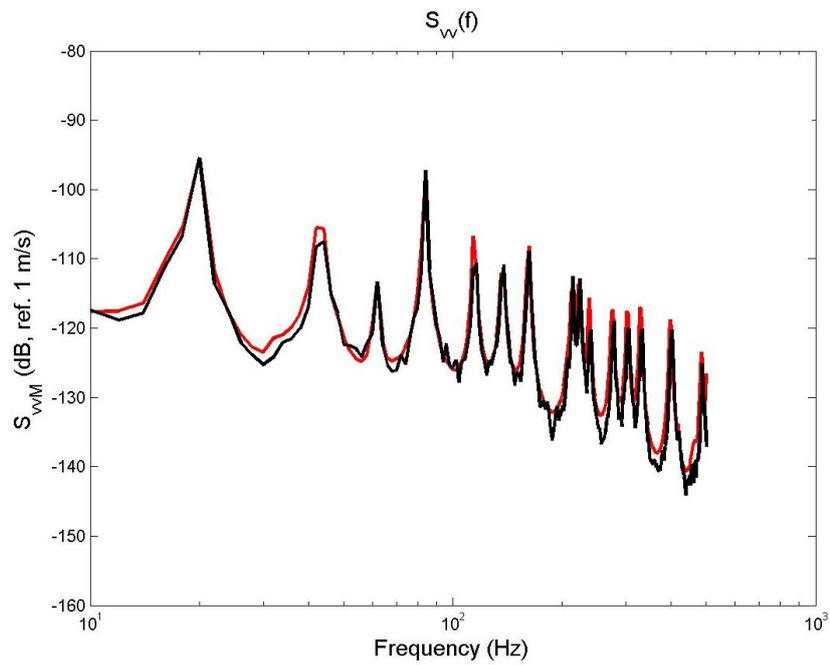

Figure 6. Comparison of the results of method 1 (red) and method 2 (black).

## 3.4 *The wavenumber method*

The third method is based on a formulation in the wavenumber space of Eq. (6). Let us consider the space Fourier transform of the wall pressure spectrum, $\phi_{pp}^{TBL}(k,\omega)$. With our definition of the Fourier transform, it is related to the wall pressure spectrum in the physical space $S_{pp}^{TBL}(\tilde{x}-\tilde{\tilde{x}},\omega)$ by

$$S_{pp}^{TBL}(\tilde{x}-\tilde{\tilde{x}},\omega) = \frac{1}{4\pi^2}\int_{-\infty}^{+\infty}\phi_{pp}^{TBL}(k,\omega)e^{jk(\tilde{x}-\tilde{\tilde{x}})}dk. \tag{17}$$

Introducing Eq. (17) in Eq. (6) gives

$$S_{vv}(x,\omega) = \frac{1}{4\pi^2}\int_{-\infty}^{+\infty}\phi_{pp}^{TBL}(k,\omega)\left|H_v(x,k,\omega)\right|^2 dk, \tag{18}$$

with

$$H_v(x,k,\omega) = \int_{S_p} H_v(x,\tilde{x},\omega)e^{jk\tilde{x}}d\tilde{x}. \tag{19}$$

$H_v(x,k,\omega)$ is generally called the sensitivity function [22]. The interpretation of Eq. (19) indicates that this quantity corresponds to the velocity at point *x* when the panel is excited by an unit wall plane wave with wavevector *k* (i.e. by a wall pressure field $e^{jkx}$, $\forall x \in S_p$).

We emphasize that theses wall plane waves can be generated by travelling acoustic plane waves only for wavenumbers *k* inside the acoustic domain (i.e. $|k| \leq k_0$, with $k_0$, the acoustic wavenumber). For wavenumbers inside the subsonic domain (i.e. $|k| > k_0$), the acoustic plane waves are evanescent and it is then more complex to generate them physically. The Source Scanning Technique proposed in [23] is one solution. From a numerical point of view, this problem disappears. The pressure field of this excitation can be directly applied as the panel loading. When using a numerical vibroacoustic model (like FEM, BEM, etc), it is however necessary to check that the spatial discretization of the model allows to represent the spatial variation of this pressure field.

The third method proposed on this paper is based on a truncation and a regular discretization of the wavenumber space *k*. We suppose that the discrete space is composed of *I* points which are noted $k_i$, $\forall i \in [\![1,I]\!]$. The ASD of the velocity at point *x* can then be approximated by

$$S_{vv}(x,\omega) \approx \frac{1}{4\pi^2}\sum_{i=1}^{I}\phi_{pp}^{TBL}(k_i,\omega)\left|H_v(x,k_i,\omega)\right|^2 \Delta k_i, \tag{20}$$

where $\Delta k_i$ is the elementary surface in the wavenumber domain attributed to the discrete wavenumber $k_i$.

The truncation and the discretisation of the wavenumber space should be done carefully in order to avoid the loss of information:

- the wavenumber resolutions in the two directions should be defined such as to permit a correct representation of the spatial variations in the wavenumber space of the sensitivity function and the wall pressure spectrum. For the present test case, an analytical calculation of the sensitivity function for an invacuo panel gives us an order of magnitude of these spatial variations (inversely proportional to the panel lengths). The wavenumber resolutions are fixed to 1 rad/m, independently of the frequency. For a more complex panel, a trial and error process would be necessary to fix these parameters;

- the cut-off wavenumbers in the two directions should be defined such as the main contributions of the integrant of (18) are well taken into account. This point is illustrated on Fig. 7 for the present test case at a given frequency. The highest values of the sensitivity functions are obtained for wavenumbers close to $k_f^{wet}$, the natural flexural wavenumber of an equivalent infinite plate taking the fluid added mass effect into account. On another hand, the wall pressure spectrum exhibits the highest values for wavenumbers close to the convective wavenumber, $k_c$. In theory, the cut-off wavenumber should be defined in the streamwise direction by $\bar{k}_x = \kappa_x \max\left[k_f^{wet}, k_c\right]$ and in the crosswise direction by $\bar{k}_y = \kappa_y k_f^{wet}$ where $\kappa_x$ and $\kappa_y$ are margin coefficient. As the considered frequency is well above the hydrodynamic coincidence frequency, we have $k_f^{wet} << k_c$ and $\bar{k}_x = \kappa_x k_c$. This last criterion can lead to huge computing costs (because the spatial discretisation of the vibroacoustic model should be able to describe the "small" wavelength $2\pi/\bar{k}_x$). However, it is well known [24] that in many cases, the structure plays a role of filtering of the excitation which is dominant. This is illustrated on Fig. 7c where the product of the sensitivity function with the wall pressure spectrum (i.e. integrant of Eq. (18)) has been plotted. It can be observed that the contribution of the convective domain is negligible. Then, for this case, the cut-off wavenumber in the streamwise direction can be reduced to $\bar{k}_x = \kappa_x k_f^{wet}$. This permits to save huge computing times. It should be emphasized that this restriction is not always valid. In particular, it depends on the frequency (compared to the hydrodynamic frequency), on the considered wall pressure model (see [25]), and the boundary conditions of the panel (see [26]). Here again, a trial and error process could be necessary at certain frequencies to fix the cut-off wavenumber;

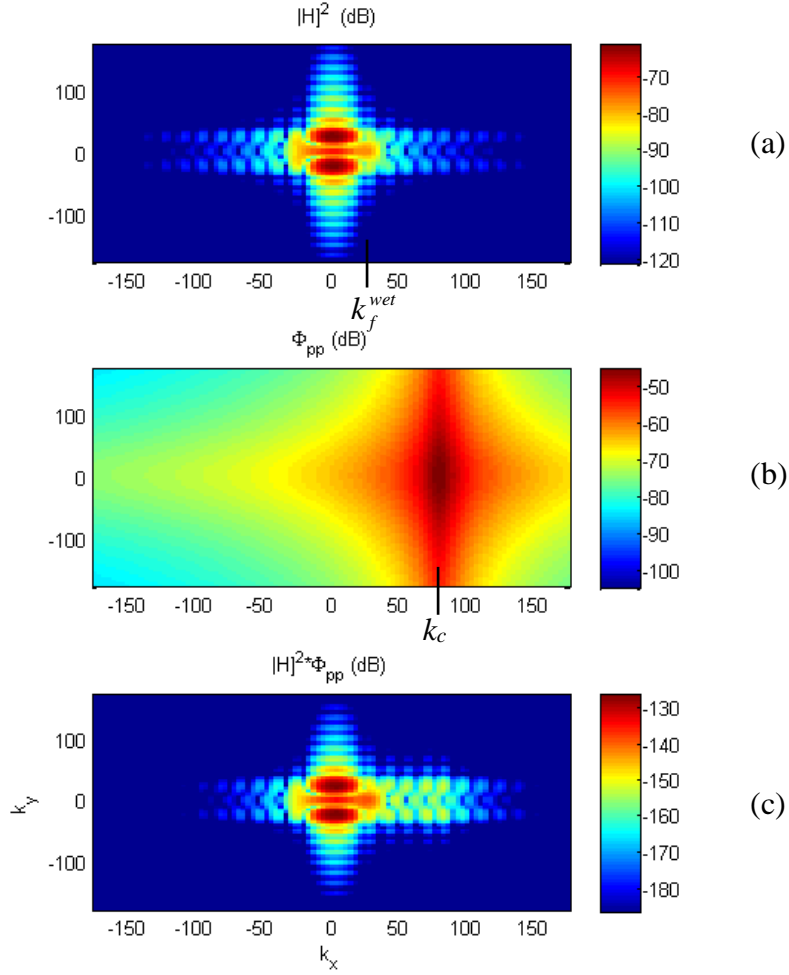

Figure 7. Different quantities in the wavenumber space $k=(k_x,k_y)$: (a), the sensitivity function at point $x$, $H_v(x,k,\omega)$; (b), the wall pressure spectrum, $\phi_{pp}^{TBL}(k,\omega)$ ; (c), the product between the sensitivity function and the wall pressure spectrum, $\phi_{pp}^{TBL}(k,\omega)|H_v(x,k,\omega)|^2$. Results presented at 100 Hz.

With this approach, the number of load cases corresponds to the number of wall plane waves considered for the calculation of the sensitivity functions, $H_v(x,k,\omega)$.

The present method respecting the previous criteria for the wavenumber resolutions and the cut-off wavenumbers is compared with the spatial method in Fig. 8. We can observe that the spatial method gives results slightly higher than the wavenumber method (excepted for the first peaks). This can be attributed to the spatial resolutions of the first method (i.e. $\delta=\lambda_c/4$) which is not sufficiently small to ensure a full convergence of the method.

Contrary to the spatial and Cholesky methods, the wavenumber method allows us to obtain results up to 1 kHz. This is mainly due to the fact that the convective ridge which can be supposed negligible is not described with this method when using appropriate cut-off wavenumbers.

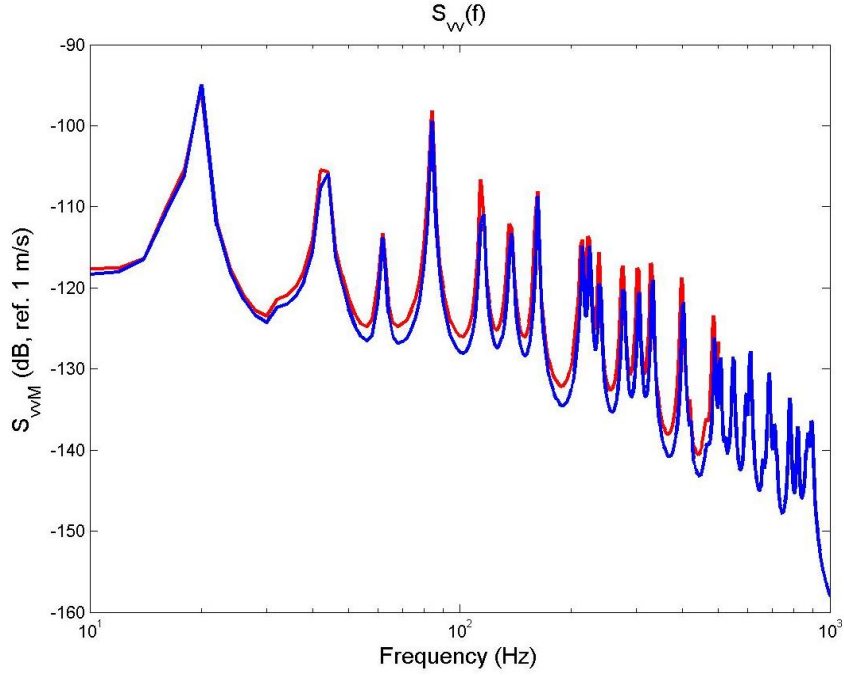

Figure 8. Velocity ASD at point $x$=(0.05, 0.18).
Comparison of the results of method 1 (red) and method 3 (blue).

### 3.5  *The reciprocity method*

This fourth approach has been proposed in [11] for predicting the noise radiated by stiffened structures excited by TBL. It is based on a reciprocity principle which gives a second interpretation of the sensitivity functions. Indeed, the Lyamshev reciprocity principle [27] for vibro-acoustic problems indicates that the ratio of the normal velocity of the plate at point $x$ over the applied normal force at point $\tilde{x}$ is equal to the ratio of the normal velocity of the plate at point $\tilde{x}$ over the normal force applied at point $x$. With the present notation, we can write:

$$H_v(x, \tilde{x}, \omega) = H_v(\tilde{x}, x, \omega). \tag{21}$$

This expression can be injected in the definition of the sensitivity function (i.e. Eq. (19)) that allows us writing

$$H_v(x, -k, \omega) = \int_{S_p} H_v(\tilde{x}, x, \omega) e^{-jk\tilde{x}} d\tilde{x}. \tag{22}$$

One recall that $H_v(\tilde{x}, x, \omega)$ represents the velocity response at point $\tilde{x}$ when the panel is excited at point $x$. Then, $H_v(x, -k, \omega)$ can be interpreted as the spatial Fourier transform of the velocity response of the panel excited at point $x$. Consequently, the power spectrum density of the velocity of the plate at point $x$ excited by the TBL can be calculated with Eq. (20) on the basis of the response of the plate excited by a normal force at point $x$ and expressed in the wavenumber space by a discrete spatial Fourier transform. That is to say that the plate response at a given point due to TBL can be estimated from the vibratory field of the plate excited by a point force at this same point.

We can emphasize that this technique remains available even if the point of observation is into the fluid domain (for example for dealing with transmission loss problem). In this case, the radiated pressure at a point $z$ by the TBL-excited panel would be estimated from the velocity field of the panel excited by an acoustic monopole located at point $z$ and having unit volume flow rate [11].

The main advantage of this approach is that the number of load case is very small in general because it corresponds to the number of receiving points for which the response to the TBL excitation should be estimated. As this excitation is spatially distributed, it is generally not necessary to consider a large number of receiving points, the stochastic vibratory field being relatively homogeneous.

We compare the sensitivity functions obtained with these two interpretations on Fig. 9. Of course, the results are very similar. The wavenumber resolutions differ as the ones of the reciprocity method depends directly on the panel dimensions (as a consequence of the discrete spatial Fourier transform). The comparison of the wavenumber and reciprocity methods on Fig. 10 shows a good agreement. The slight differences can be attributed to the different wavenumber resolutions.

We can emphasize that this method requires few load cases but it requires evaluating the spatial distributions of the vibratory field in order to perform the spatial Fourier transform. When the vibratory field is evaluated from a numerical model (like the PTF approach used in the present paper), this task can be relatively time consuming and can reduce the efficiency of this approach. At its origin, this approach has been developed for dealing with stiffened structures like plate or shell stiffened in 1 direction or in 2 orthogonal directions [11]. For these cases, it is possible to calculate analytically the sensitivity functions thanks to the reciprocity principle described in this section. The computing times are then very short.

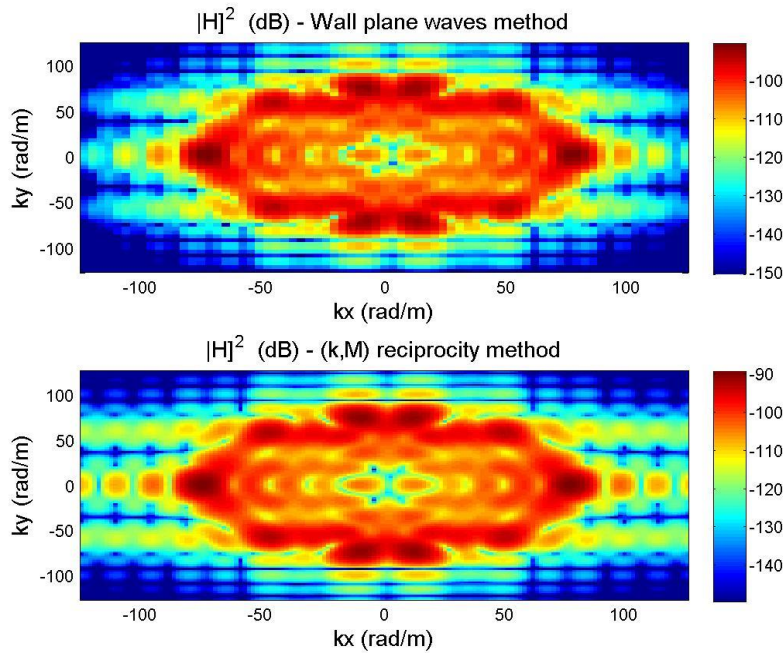

Figure 9. Sensitivity function at point $x=(0.05, 0.18)$ in function of the wavevector $k=\left(k_x,k_y\right)$. Results at 1 kHz. Two calculations: upper, with Eq. (19) (i.e. direct interpretation); lower, with Eq. (22) (i.e. using the reciprocity principle).

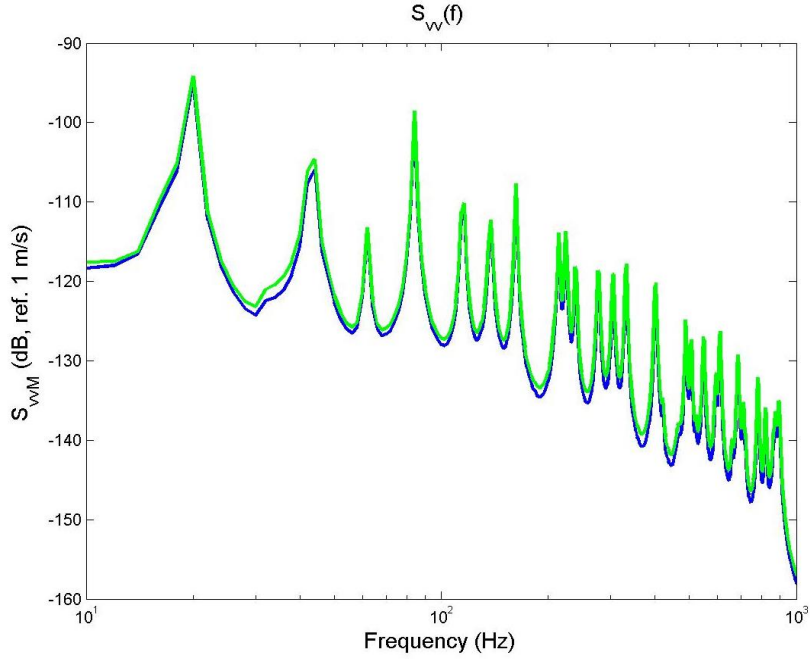

Figure 10. Velocity ASD at point $x$=(0.05, 0.18).
Comparison of the results of method 3 (blue) and method 4 (green).

## 3.6 *Method based on the sampling of uncorrelated wall plane waves*

The last of the five methods presented in this paper has been presented recently in [28]. It has some similarities with the method 2 (Sec. 3.3). But, contrary to the latter, it does not require a Cholesky decomposition (which can be time consuming).

Basically, it consists in rewriting Eq. (20) in the following form:

$$S_{vv}(x,\omega) \approx \sum_{i=1}^{I} S_{A_i A_i}(\omega) \left| H_v(x,k_i,\omega) \right|^2, \qquad (23)$$

with

$$S_{A_i A_i}(\omega) = \frac{\phi_{pp}^{TBL}(k_i,\omega) \Delta k_i}{4\pi^2}. \qquad (24)$$

This expression can be interpreted as the panel response to a set of uncorrelated wall plane waves of stochastic amplitudes $A_i$, $i \in [\![1,I]\!]$. $S_{A_i A_i}(\omega)$ represents the ASD of the amplitude of the $i^{th}$ waves. These wall plane waves are uncorrelated because Eq. (23) corresponds to the case where $S_{A_i A_j}(\omega) = 0$, $\forall i \neq j$ (see [23] for details).

This interpretation is similar to the one generally considered for describing an acoustic diffuse field: a set of uncorrelated acoustic waves of equiprobable incident angles and equal intensities. In the present case with the TBL excitation, the waves are not limited to the acoustic domain and their amplitudes are not constant; they depend on the wall pressure fluctuations, $\phi_{pp}^{TBL}(k_i,\omega)$ from Eq. (24).

From this interpretation, we can define the wall pressure field of the $k^{th}$ realization, $p^k(x_i, \omega)$ by,

$$p^k(x_i, \omega) = \sum_{i=1}^{I} \sqrt{S_{A_i A_i}(\omega)} e^{j\varphi_i^k} e^{jkx_i}, \qquad (25)$$

where $\varphi_i^k$, $\forall i \in [\![1, I]\!]$, are random phase values uniformly distributed in $[0, 2\pi]$.

As for the method 2, the panel velocity at point $x$ when the panel is excited by the pressure field, $p^k(\omega)$ is then estimated by using the vibroacoustic model. This process is repeated for a given number of realizations, $K$; and, finally, the ASD of the velocity at point $x$ is estimated by an ensemble average of the velocity responses, $v^k(x, \omega)$ (see Eq. (16)).

In Fig. 11, the velocity responses of 20 realizations are plotted in grey and the ensemble average over these 20 realizations is plotted in black. We can observe that this figure is similar to Fig. 5 related to the method 2, excepted that the calculation is achieved up to 1 kHz. The advantages of the present method compared to the method 2 are: (a), it is not necessary to use a Cholesky decomposition to define the wall pressure field of each realization. Eq. (25) with Eq. (24) can be applied directly from the wall pressure spectrum expressed in the frequency-wavenumber space (with a Corcos or a Chase model for example); (b), the use of adapted cut-off wavenumbers permits to neglect easily the effect of the convective ridge and then to save computing time.

The good agreement between the results of method 3 and 5 on Fig. 12 allows us validating the present approach.

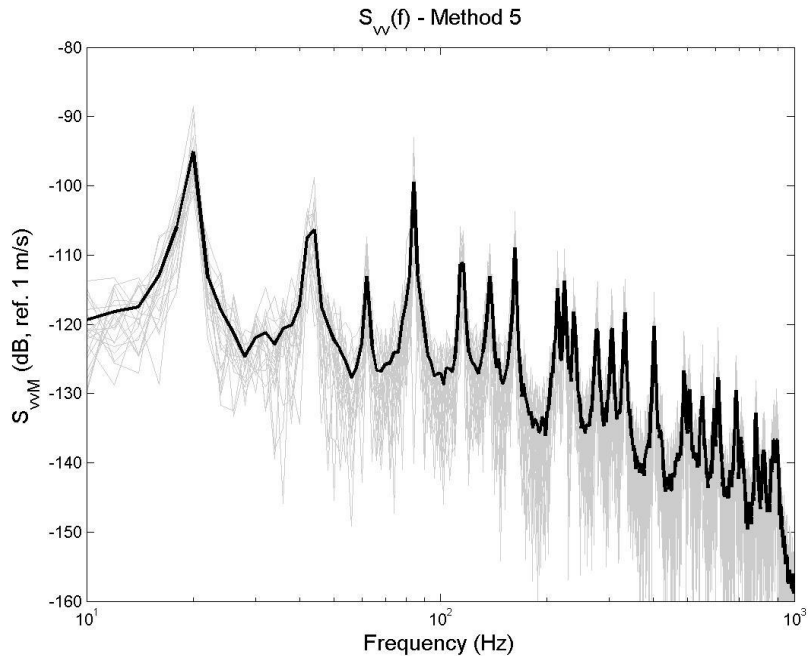

Figure 11. Velocity ASD at point $x$=(0.05, 0.18).
Grey, Results of 20 realizations; Black, Average over the 20 realizations.

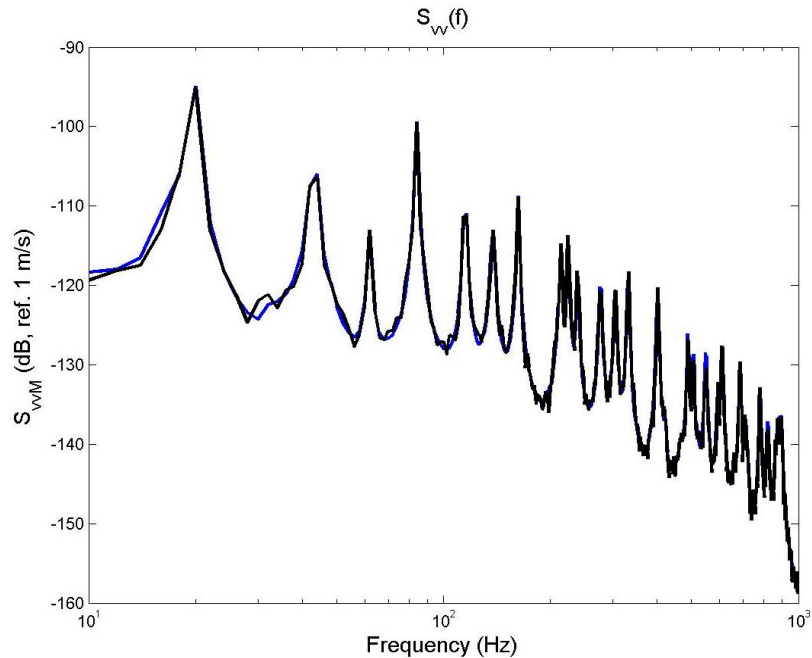

Figure 12. Comparison of the results of method 3 (blue) and method 5 (black).

## 3.7 *Synthesis*

Five methods for coupling a TBL wall pressure model with a deterministic vibro-acoustic model have been presented:

- The first two methods are adapted for a wall pressure spectrum expressed in the physical space (like given by the Corcos model); the spatial resolution criterion which permits to describe correctly the convective ridge requires a very fine discretization of the panel surface and it can limit these methods to low frequencies and/or to small panels. We have noticed that if this criterion is not respected, these two methods overestimate significantly the vibratory field;

- The three other methods are adapted for a wall pressure spectrum expressed in the wavenumber space. In some situations (e.g. for frequencies well above the hydrodynamic frequency), the effect of the convective ridge can be neglected which enables to reduce the cut-off wavenumber (with a criterion based on the panel characteristics and not on the TBL ones). It permits to save computing time. Thanks to that, a higher frequency range has been reached by comparison with the spatial methods. It should however be emphasized that if the wall pressure spectrum in the physical space was filtered by a low pass filter in order to suppress the convective peak corresponding to the small spatial separations, the spatial methods would certainly have a similar efficiency than the wavenumber methods.

As an indication, we give on Tab. 2 the number of load cases and the CPU time per frequency observed on the present test case for the 5 methods. We should emphasize that these computing times do not represent strictly the efficiency of each method; they depend strongly on the calculation algorithm (that we try to optimize), the computing environment (MATLAB for us), the management of the input/output of the vibroacoustic code (use of PTF in-house code). Anyway, the method based on the realizations of uncorrelated wall plane waves gave us smaller computing time. The reciprocity method is the one which necessitates the lowest number of load case. It could be the most efficient if the management of the input/output with the vibroacoustic code would not be optimal.

The presented results and the discussion focus on the vibratory response of the panel. Of course, all the methods described in this paper can be used to evaluate the radiated pressure from the panel excited by TBL (for dealing with transmission loss problem for example). Moreover, they can be applied to more complex structures than the rectangular thin plate considered for illustration.

| Method | Spatial | Cholesky | Wavenumb. | Reciprocity | Uncorrelated waves |
|---|---|---|---|---|---|
| Number of load cases | 27300 | 20 | 10000 | 1 | 20 |
| CPU time / frequency (s) | 5.8 | 16.4 | 2 .5 | 2.3 | 1.9 |

Table 2. Synthesis of the number of load cases and the computing times.

# 4. High frequency modelling

## 4.1 *Statistical Energy Analysis*

Statistical energy analysis (SEA) allows the vibro-acoustic behaviour of complex structures in high-frequency range to be predicted [7, 29]. The method is based on a fundamental relationship relating the power flow exchanged by two-coupled subsystems to their total subsystem energies by the coupling loss factor (CLF).

Basically, SEA consists in decomposing the global subsystem in different subsystems as illustrated in Fig. 13 for a Sonar self noise issue on a ship. This substructuring should be done in order to fulfil several conditions [30-33]. In particular, each subsystem should exhibit several (many) modes in the frequency band of interest and the couplings between subsystems should be weak [31]. For the case presented in Fig. 13, the coupling between the dome and the Sonar cavity filled of water may be a problem because it does not respect strictly the weak coupling assumption [34]. In this case, SEA can be seen as a first approximate model which is valuable for practical studies [35].

In a second step, SEA consists in writing the power balance for stationary motion in each subsystem using the fundamental SEA relation to evaluate the power flows. It produces a linear equation system where the unknowns are the total energies of subsystems. Then, the difficulty in applying SEA is not due to solving complicated equations, but in evaluating the SEA input parameters such as the damping loss factors, the coupling loss factors [36] and the injected power [37-39].

In this section, we focus the discussion on the evaluation of the injected power when the SEA subsystem is excited by a TBL. For the illustration case of Fig. 13, it consists in estimating the injected power by the turbulent flow in three subsystems (i.e. the Sonar dome and the two parts of the hull) for each frequency band (typically, third octave bands).

We suppose that the TBL parameters characterizing the turbulent flow have been obtained from a hydrodynamic code and an appropriate model allows us describing the spectrum of the wall pressure fluctuations.

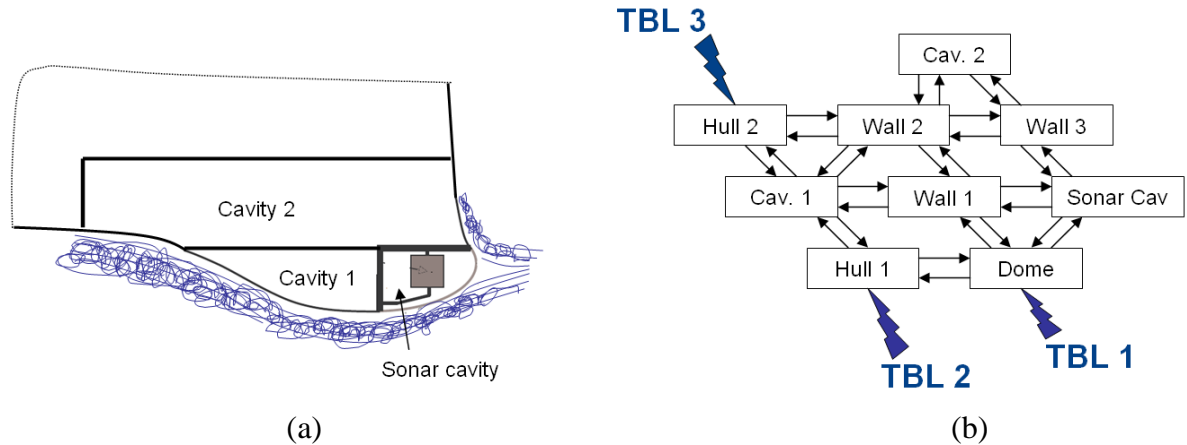

Figure 13. (a), Illustration of the self noise issue on the bow of an Anti-Submarine Warfare surface ship; (b), SEA model describing the energy sharing between subsystems.

Before going into the details about the injected power calculation, it is necessary to make a break on two points:

- First, it should be remembered that the injected power depends not only on the excitation, but also on the receiving structure. If the structure was infinitely rigid, the injected power would be null. In the case of SEA model for which the global system has been decomposed in weakly coupled subsystems, one can suppose that the injected power in a given subsystem can be evaluated by neglecting the couplings with the others subsystems. This assumption is valid if the weak coupling condition is well respected with the others subsystems;

- The calculation of the injected power should be performed in the framework of SEA hypothesis. That is to say that the frequency is relatively high, the excited structure presents many modes, and the different SEA quantities (especially the injected power) are time-averaged for a considered frequency band. Under these hypotheses, it may be reasonable to evaluate the injected power in a complex structure from the one in an equivalent academic structure [7]. Generally, this latter is a rectangular thin plate. It could be surprising at the first sight to "replace" a complex structure like the Sonar dome by a thin plate. However, it is well-know that in the high frequencies, the effects of curvature of the dome and of the boundary conditions on its vibratory behaviour can be neglected. If the dome is made of an isotropic material and of constant thickness, a thin plate can then be reasonably considered to evaluate SEA parameters like the modal density or the injected power. For more complex structures like the stiffness hull of the ship, the thin plate alone is probably not sufficient to represent correctly the behaviour at high frequencies. In particular, the propagation of the Bloch-Floquet waves due to the periodic stiffeners would not be described. This aspect of approximation is part of the difficulties in applying SEA to manufactured structures and is also part of the expertise of the SEA specialists.

Anyway, in this paper, we have decided to focus our attention on the estimation of the power injected by the TBL in an equivalent thin plate.

## 4.2 Estimation of the time-average injected power

Let us consider a thin plate subjected to a TBL excitation. The plate is made of an isotropic elastic material and has a constant thickness. $M, D, h, \eta_s$ are, respectively, the mass per unit area, the flexural rigidity, the thickness, and the damping loss factor of the plate. The TBL is fully developed, stationary and homogeneous. We consider a frequency band of angular bandwidth $\Delta\Omega$ and of central angular frequency $\Omega$ which is well above the hydrodynamic coincidence angular frequency.

The energy balance equation consists in writing that the injected power by the TBL is dissipated by the plate. The time average of the injected power in the considered frequency band, $\Pi_{inj}$ can then be evaluated from

$$\Pi_{inj} \approx \Omega \eta_s M <V^2>, \tag{26}$$

where $<V^2>$ is the time and space average of the quadratic velocity of the plate.

In the high frequencies, the shape of the plate and the type of boundary conditions do not influence the SEA parameters [7]. Then, a rectangular simply-supported plate was considered in [24, 37] for evaluating $<V^2>$ from a modal expansion. An alternative consists in considering an infinite plate (which is excited by a 'fictive' homogeneous CLT) and in evaluating $<V^2>$ for a given area, $S_p$ of the plate. This "equivalence" of vibratory behaviour in the high frequencies between a finite structure and an infinite one is often used in SEA. For its simplicity in the mathematical developments, we adopt it in the present paper.

As the infinite plate is theoretically excited by a homogeneous CLT, the vibratory field is assumed to be spatially homogeneous. The ASD of the velocity at a given point $x$ is independent of the point position:

$$S_{vv}(x,\omega) = S_{vv}(\omega), \ \forall x \in S_p. \tag{27}$$

The time and space average of the quadratic velocity of the plate is obtained from

$$<V^2> = \frac{1}{2\pi} \int_{\Omega-\Delta\Omega/2}^{\Omega+\Delta\Omega/2} S_{vv}(\omega) d\omega. \tag{28}$$

The wavenumber formulation of Sec. 3.4 allows us writing the ASD of the velocity

$$S_{vv}(\omega) = \frac{1}{4\pi^2} \int_{-\infty}^{\infty} \int_{-\infty}^{\infty} \phi_{pp}^{TBL}(k_x, k_y, \omega) |H_v(k_x, k_y, \omega)|^2 dk_x dk_y, \tag{29}$$

where $k_x, k_y$ are wavenumbers in the streamwise and spanwise directions, respectively.

The sensitivity functions, $H_v(k_x, k_y, \omega)$ can be calculated using the reciprocity principle described in Sec. 3.5. It corresponds to the transversal velocities of the plate expressed in the wavenumber space when the plate is excited by a normal point force at an arbitrary point. (We chose the coordinate origin for convenience.) Considering the Kirchhoff-Love's dynamic plate equation, we obtain:

$$H_v(k_x, k_y, \omega) = \frac{j\omega}{D\left[(1+j\eta_s)(k_x^2+k_y^2)^2 - k_f^4\right]}, \tag{30}$$

with $k_f = \omega^{1/2}\left(\frac{M}{D}\right)^{1/4}$, the natural flexural wavenumber of the plate.

We can notice that these sensitivity functions have the most important magnitudes for wavenumbers close to the flexural wavenumber (i.e. when $\sqrt{k_x^2 + k_y^2} \approx k_f$) and their magnitudes decrease quickly when the wavenumbers deviate from these values (see example Fig. 14a). On the contrary, the wall pressure spectrum varies relatively slowly in the subconvective wavenumber domain (see Fig. 14b). Then, the more significant contributions of the integrand of (29) correspond to the wavenumbers close to the flexural wavenumber (taking account that the convective ridge can be neglected seeing that the frequency band of interest is well above the hydrodynamic coincidence frequency). Supposing moreover that the wall pressure spectrum is relatively flat for wavenumbers close to the flexural wavenumber, we can write:

$$S_{vv}(\omega) \approx \frac{\phi_{pp}^{TBL}(k_f, 0, \omega)}{4\pi^2} \int_{-\infty}^{\infty}\int_{-\infty}^{\infty} |H_v(k_x, k_y, \omega)|^2 \, dk_x dk_y, \tag{31}$$

This approximation is illustrated in Fig. 14c by comparing $\phi_{pp}^{TBL}(k_x, k_y, \omega)|H_v(k_x, k_y, \omega)|^2$ (full line) with $\phi_{pp}^{TBL}(k_f, 0, \omega)|H_v(k_x, k_y, \omega)|^2$ (dashed line); as discussed in Sec. 3.4, it should be emphasized that this restriction is not always valid. The validity of this approximation depends on the frequency (compared to the hydrodynamic frequency), on the considered wall pressure model, and the boundary conditions of the panel. In the 'high' frequency, it is generally well respected.

The integral of this expression can be approximated by

$$\int_{-\infty}^{\infty}\int_{-\infty}^{\infty} |H_v(k_x, k_y, \omega)|^2 \, dk_x dk_y \approx \frac{\omega^2}{8D^2\eta_s k_f^6}. \tag{32}$$

Introducing Eqs. (28, 31, 32) in Eq. (26) and supposing that $\Delta\Omega \ll \Omega$ (and then $\omega \approx \Omega$), we obtain an estimation of the injected power by the TBL in the plate:

$$\Pi_{inj} \approx \frac{S}{4\sqrt{MD}} \phi_{pp}^{TBL}(k_f, 0, \Omega)\Delta\Omega. \tag{33}$$

An expression which differs only by a $(2\pi)^3$ factor was obtained in [37] considering a simply supported plate and a modal calculation. This factor is only due to the difference of definition of the space-time Fourier transforms between the two papers.

We notice that this power is independent from the plate damping. This may have consequences for vibration and noise control. As it can be expected, it is also proportional to the area excited by the turbulent flow.

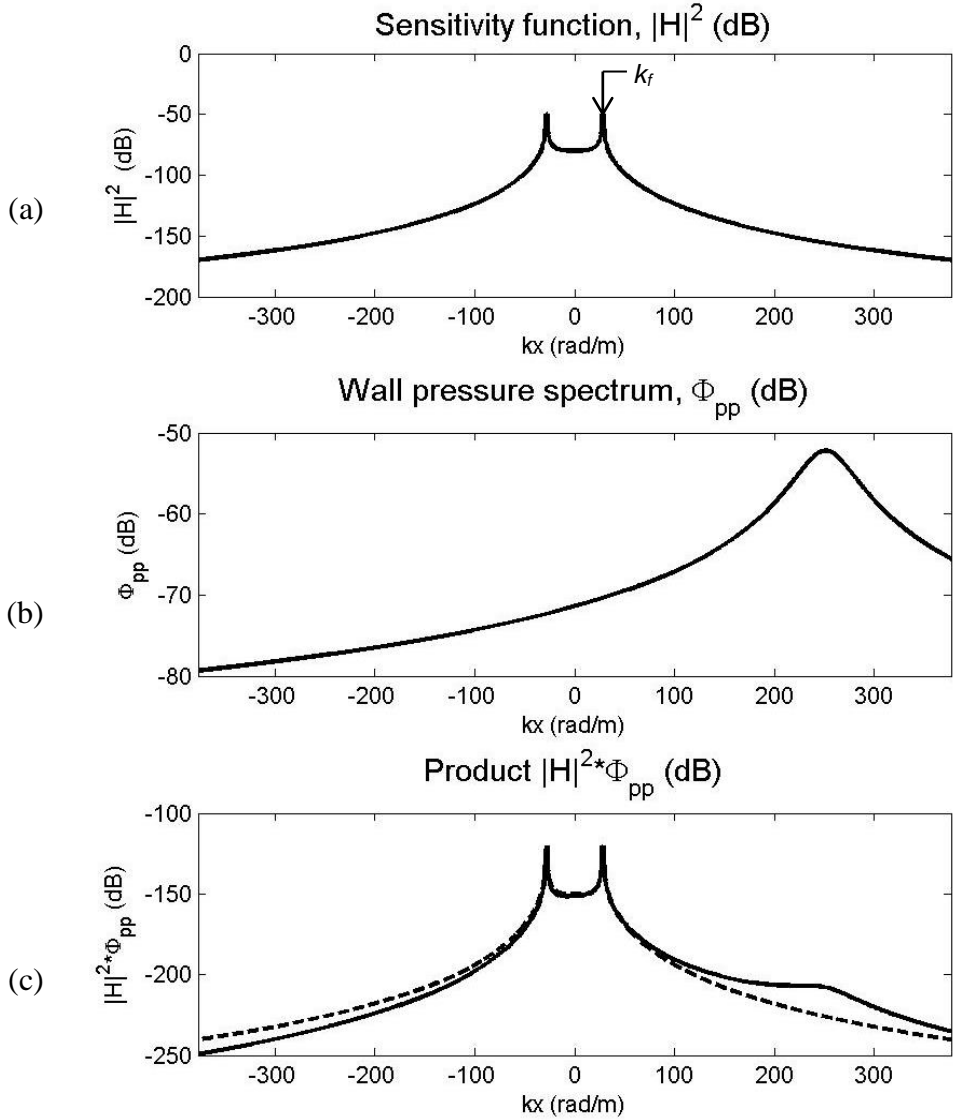

Figure 14. (a), Sensitivity function for an infinite steel 1mm-thick plate; (b), Corcos wall pressure spectrum; (c), Product between the sensitivity function and the wall pressure spectrum: full, without approximation; dotted, with the approximation used in Eq. (31). Results at 200 Hz for $k_y=0$ rad/m.

Different approximations have been made to obtain this formula. In particular, the frequencies should be well higher than the hydrodynamic coincidence frequency and the wall pressure spectrum should be considered relatively flat in the wavenumber region concerned by the plate characteristic wavenumbers (i.e. flexural wavenumber). Comparison in [37] with an "exact calculation" for a present test case and considering the Corcos model showed that the discrepancies were very small in the frequency domain for which the SEA can be applied. For aeronautical application, the calculation of the injected power proposed in Ref. [38-39] can be more accurate for frequencies lower than or close to the aerodynamic coincidence frequency.

Expression (33) has been obtained considering a flat plate and it may give a fair approximation of the injected power in subsystems composed of sheets having a high radius of curvature, roughly constant thickness and made of isotropic material. For stiffened structures like the ship hull, it could be seen as a first approximation. A more accurate prediction could be obtained by considering the sensitivity functions of a periodically stiffened plate [11]. In this way, it would be difficult to obtain an analytical expression of the injected power but a numerical process could be developed. SEA results could be compared to the approach proposed in [40-42] to estimating broadband levels of acoustic power radiated due to rib/panel interaction under TBL-like excitation.

## 4.3 *A methodology for taking into account the spatial variation of the TBL parameters*

Hydrodynamic codes [37] permit to estimate spatial variations of the TBL parameters due to static pressure gradients or development of the TBL. An illustration is given on Fig. 15 for the bow of an Anti-Submarine Warfare surface ship. The TBL parameters can then vary on the surface of a given SEA subsystem (for example the Sonar dome of the ship). This can be an issue for evaluating the SEA injected power. However, if these variations are relatively slow compared to the wavelengths of the flexural motions, a numerical process taking these variations into account can be proposed. Indeed, we have noticed previously that the relation (33) has been obtained independently of the boundary conditions of the panel and it remains valid as long as many wavelengths are contained along each edge of the panel. Then, it can be use to evaluate the injected power in a part of a subsystem for which TBL parameters does not vary significantly.

The process consists in dividing the subsystem surface (excited by TBL) in *K* patches having roughly constant TBL parameters. For each patch $k$ of surface $S^k$, we can evaluate the injected power per unit area, $\pi_{inj}^k$:

$$\pi_{inj}^k = \frac{1}{4\sqrt{MD}} \phi_{pp,k}^{TBL}(k_f, 0, \Omega) \Delta\Omega, \quad k \in [\![1, K]\!], \tag{34}$$

where $\phi_{pp,k}^{TBL}$ is the wall pressure spectrum depending on the TBL parameters on the $k^{th}$ patch.

An approximation of the injected power can then be obtained from

$$\Pi_{inj} \approx \sum_{k=1}^{K} \pi_{inj}^k S^k. \tag{35}$$

An illustration of this process is given on Fig. 16. The injected power in the Sonar dome is evaluated by integrating over the dome surface the injected power by unit area. The latter has been calculated from the parameters of Fig. 15 and it exhibits significant spatial variations. This highlights the importance to taken the TBL parameters into account.

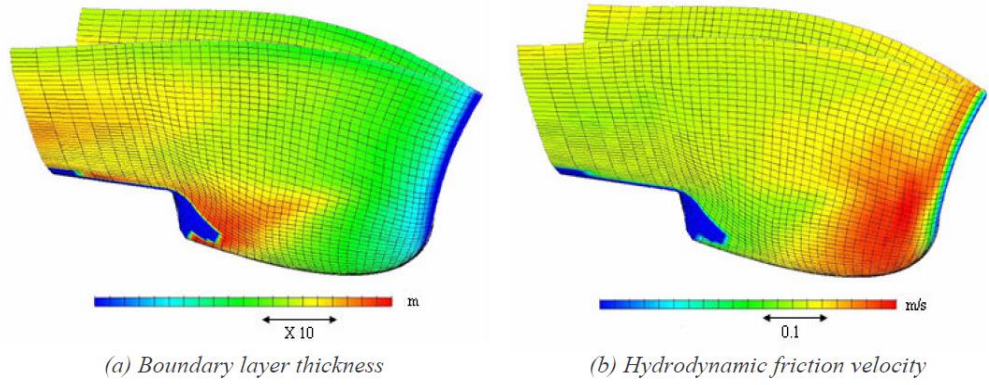

Figure 15. Illustration of hydrodynamic calculation of the TBL parameters for the front of a ship [37]: (a), Boundary layer thickness; (b), Hydrodynamic friction velocity.

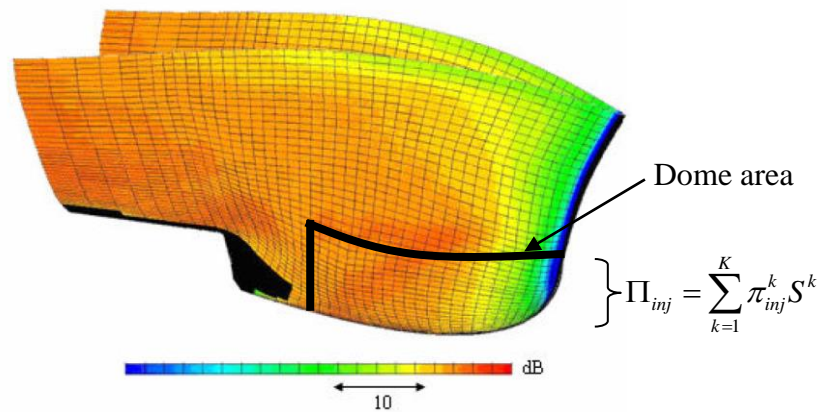

$$\Pi_{inj} = \sum_{k=1}^{K} \pi_{inj}^{k} S^{k}$$

Figure 16. Values of the injected power by unit area obtained from the TBL parameters of Fig. 15 [37].

## 5. Conclusions

This paper is focused on the coupling between TBL excitations and vibro-acoustic models. In the low frequency range, different techniques have been presented to make the relationship between the stochastic excitation and the deterministic model. The efficiency of these techniques in terms of computation time depends on various parameters such that the choice of the wall pressure models (in spatial or wavenumber form), the values of the physical parameters of the considered case, the efficiency of the vibro-acoustic code (in particular, its ability to manage multi-load cases), etc. For the marine test considered in this paper, the method consisting in the realizations of uncorrelated wall plane waves was found to be the fastest one. This method is easy to implement and it requires a small number of vibro-acoustic calculations (i.e. the number of load cases is equal to the number of realization). These methods offer a large possibility for coupling the wall pressure spectrum of the CLT excitation with the transfer functions describing the vibro-acoustic behaviour of the considered structure. In the future a more detailed study of the influence of the spatial variations of the TBL parameters should be undertaken.

In the high frequency range, a formulation of the injected power in a SEA subsystem subjected to a TBL excitation has been proposed as a function of the wall pressure spectrum expressed in the frequency-wavenumber space. It has been obtained considering an infinite flat plate and several assumptions which are generally valid for high frequencies. Investigations should be performed in the future to extend these developments to more complex cases such as the stiffened structures frequently met in industrial applications.